\documentclass[pss]{wiley2sp} 
\usepackage{amsmath}

\begin{document}

\title{Electronic quantum optics beyond the integer quantum Hall effect}

\titlerunning{Electronic quantum optics...}

\author{%
  Dario Ferraro\textsuperscript{\Ast,\textsf{\bfseries 1}},
  Thibaut Jonckheere\textsuperscript{\textsf{\bfseries 1}}, 
  J\'er\^ome Rech\textsuperscript{\textsf{\bfseries 1}},
	Thierry Martin\textsuperscript{\textsf{\bfseries 1}}} 

\authorrunning{Dario Ferraro et al.}

\mail{e-mail
  \textsf{Dario.Ferraro@cpt.univ-mrs.fr}
	}

\institute{%
  \textsuperscript{1}\, Aix Marseille Univ, Univ Toulon, CNRS, CPT, Marseille, France}

\received{XXXX, revised XXXX, accepted XXXX} 
\published{XXXX} 

\keywords{Nanophysics, quantum noise, single electron sources, integer/spin quantum Hall effect, Andreev reflection.}

\abstract{%
%
%
%
\abstcol{%
The analog of two seminal quantum optics experiments are considered in a condensed matter setting with single electron sources injecting electronic wave packets on edge states coupled through a quantum point contact. When only one electron is injected, the measurement of noise correlations at the output of the quantum point contact corresponds to the Hanbury-Brown and Twiss setup. When two electrons are injected on opposite edges, the equivalent of the Hong-Ou-Mandel collision is achieved, exhibiting a dip as in the coincidence measurements of }{quantum optics. The Landauer-B{\"u}ttiker scattering theory is used to first review these phenomena in the integer quantum Hall effect, next, to focus on two more exotic systems: edge states of two dimensional topological insulators, where new physics emerges from time reversal symmetry and three electron collisions can be achieved; and edges states of a hybrid Hall/superconducting device, which allow to perform electron quantum optics experiments with Bogoliubov quasiparticles.}}

%
%

\maketitle   

\section{Introduction.}

Electronic quantum optics (EQO) \cite{grenier_electron_2011,bocquillon_electron_2014} aims at exploring the intimate nature of solid states systems by generating, manipulating and measuring individual electronic wave-packets (WPs) ballistically propagating in mesoscopic devices, in the same spirit as what is conventionally done for photons transmitted along wave-guides. This opens new prespectives for real time electron interferometry in the context of nanophysics.
For this purpose, on-demand single electrons and holes sources (SES) have been achieved, for instance by means of driven mesoscopic capacitors \cite{feve_-demand_2007,mahe_current_2010,buttiker_mesoscopic_1993,moskalets_quantized_2008} coupled via a quantum point contact (QPC) to the edge states of an integer quantum Hall (IQH) system,
or via properly designed Lorentzian voltage pulses \cite{dubois_integer_2013,grenier_fractionalization_2013,dubois_minimal_2013} imposed on a two dimensional electron gas. The edge states in the IQH effect, which are exempt of backscattering, play the role of such wave guides and a QPC placed downstream is equivalent to a half-silvered mirror, as it partitions electrons which are either reflected by or transmitted through the QPC. Electrons differ from photons in many aspects: they obey fermionic statistics, they are charged particles which interact among themselves and with their environment and, finally, they are always accompanied by a Fermi sea close to which electron hole pairs may be easily generated. Moreover, while in quantum optics experiments the coincidence rate is measured at the two outputs, here the noise cross-correlation signal at zero frequency is typically computed or measured \cite{bocquillon_electron_2014}.  

One of the main achievement of EQO has been the realization of the electronic equivalent of the Hanbury-Brown and Twiss \cite{hanburybrown_test_1956} (HBT) experiment where a single electron source injects electrons on a QPC, followed by the Hong-Ou-Mandel \cite{hong_measurement_1987} (HOM) experiment, where two electrons incident from two independent sources collide at the QPC. These scenarios represent fundamental tests of quantum mechanics for both photons and electrons, as they probe both their statistics and the form of the injected WPs. In the electronic HBT interferometer the Pauli principle leads to the anti-bunching between the injected electrons and the thermal ones incoming from the second channel \cite{bocquillon_electron_2012a}. In the HOM fermionic setup, when the emissions of the two colliding electrons are perfectly synchronized, one expects a suppression of the noise due again to the Pauli principle because the electrons are forced to emerge on opposite sides of the QPC. Conversely, for a long enough delay, twice the HBT signal is obtained for the noise in the HOM setup as the two sources are independent \cite{jonckheere_electron_2012}. Experiments \cite{bocquillon_coherence_2013} do validate the presence of a Pauli dip in the noise correlations, although this dip does not fall to zero at coincident injection: this is attributed to decoherence effects because of the presence of two or more interacting channels \cite{wahl_interactions_2014}.

In this paper, we wish to point out that EQO is not limited to the IQH regime. In addition to presenting the basic physics of HOM interferometry for IQH at filling factor $\nu=1$, we  explore two different situations where the paradigms of EQO are also present.  First, we  consider the situation of the quantum spin Hall (QSH) effect in a two dimensional (2D) topological insulator \cite{hasan_colloquium_2010,qi_topological_2011}, where two counter-propagating edge states carrying electrons with opposite spin appear on each side of the Hall bar \cite{ferraro_electronic_2014}. In this situation (with or without spin flip processes at the QPC) it is possible to study the interplay between Fermi statistics and time reversal symmetry (TRS) using HOM interferometry, with two or even three SES. Next, we  study the interplay between the IQH effect and superconductivity, as Andreev reflections convert an injected electron into a coherent superposition of electrons and holes -- a Bogoliubov quasiparticle -- which can subsequently be used as an input  quasiparticle for an HBT or an HOM interferometric device \cite{ferraro_nonlocal_2015}. Such Bogoliubov quasiparticles, although generated from electrons of finite energy above the Fermi sea, have recently been identified as possible realization of Majorana fermions in a condensed matter system \cite{majorana_teoria_2008,chamon_quantizing_2010,beenakker_annihilation_2014}. For simplicity, in both of these extensions, we work solely with scattering theory and do not include interactions between the edges or with the electromagnetic environment surrounding the device.      

\section{Hong-Ou-Mandel electron collisions in the integer quantum Hall effect.}

\begin{figure}[t]
\includegraphics*[width=\linewidth]{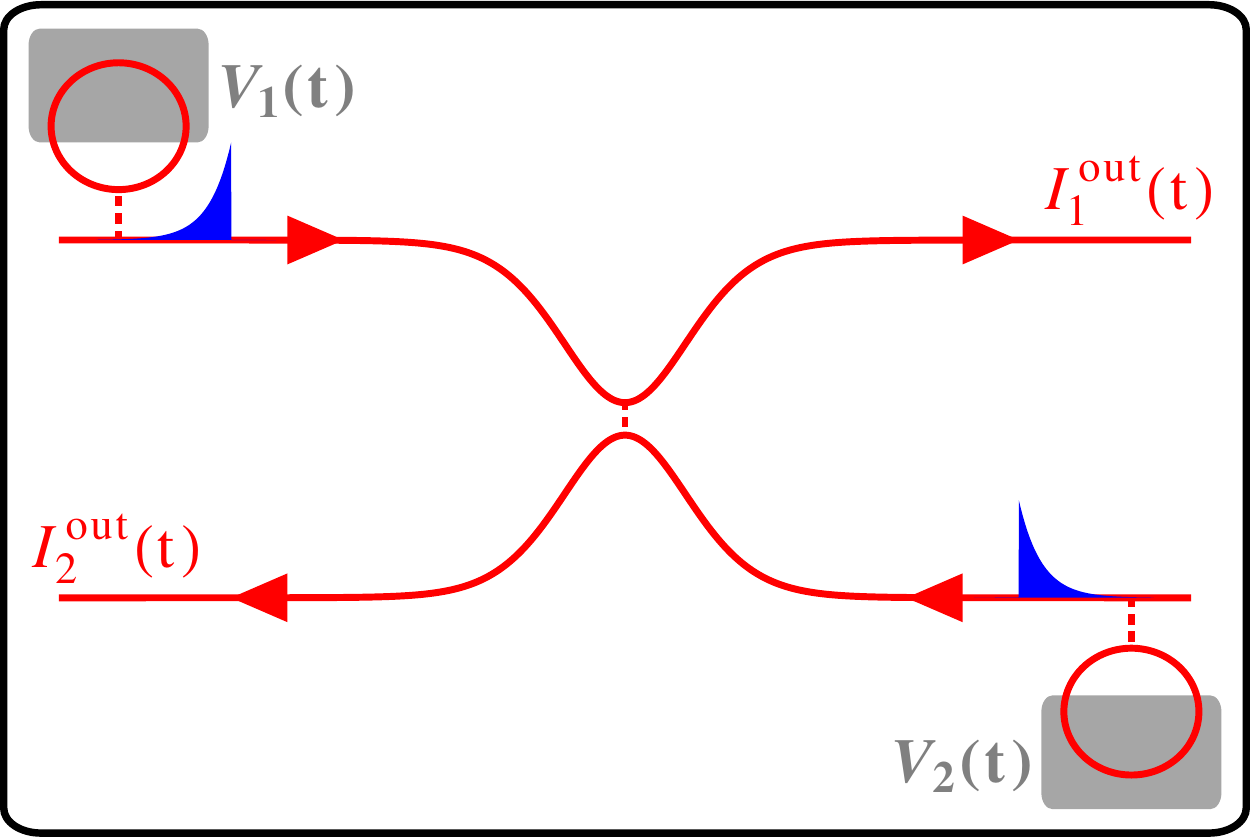}
\caption{(Color online) HOM setup in the IQH regime. Two chiral edge channels meet at a QPC. Each one is coupled to a SES in the optimal regime of emission of electrons (holes). Current cross-correlations at the two outputs are measured at zero frequency
as a function of the delay between the electron emissions. Picture taken from \cite{jonckheere_electron_2012}.}
\label{fig1}
\end{figure}

We start by considering the case of standard IQHE in the QPC geometry \cite{olkhovskaya_shot_2008} (see Fig.~\ref{fig1}). A SES injects electrons on each incoming edge state, and current cross-correlations are detected at the two outputs of the QPC, exhibiting a dependence on the delay between injections.
In experiments, the voltage drive is typically a periodic square wave, which results in the controlled emission of a regular train of single electrons and holes \cite{feve_-demand_2007,mahe_current_2010,parmentier_current_2012}.

We consider the outgoing current cross-correlations at zero frequency:

\begin{equation}
\mathcal{S}_{12}^{\,\mathrm{out}} = \int dt \, dt'  \left \langle I_1^{\,\mathrm{out}}(t) I_2^{\,\mathrm{out}}(t') \right \rangle_{c} 
\label{eq:noisestart}  
\end{equation}
where $I_1^{\,\mathrm{out}}(t)$ and $I_2^{\,\mathrm{out}}(t)$ are the currents outgoing from the QPC (see Fig.~\ref{fig1}) and where we defined the connected correlator $\langle A B \rangle_{c}= \langle A B \rangle-\langle A  \rangle\langle B \rangle$.
We can safely assume a linear dispersion for the electrons along the edge and chirality (from now on we assume the Fermi velocity $v_F=1$, reintroducing it only where needed). We compute the cross-correlations at the output of the QPC, using conventionally a x-axis pointing in the propagation direction of each edge.
A scattering matrix with transmission (reflection) probability ${\mathcal T}$ (${\mathcal R}=1-{\mathcal T}$)
characterizes the QPC. The currents outgoing from the QPC can be written in terms of the incoming field operators as
\begin{align}
I_{1}^{\, \mathrm{out}} &=  {\mathcal T} I_1 +  {\mathcal R} I_2 + i e 
           \sqrt{\mathcal{R} \mathcal{T}}(\Psi^{\dagger}_1 \Psi_2 - \Psi^{\dagger}_2 \Psi_1)  \nonumber \\ 
I_{2}^{\, \mathrm{out}} &=  {\mathcal R} I_1 + {\mathcal T} I_2 - i e 
           \sqrt{\mathcal{R} \mathcal{T}} (\Psi^{\dagger}_1 \Psi_2 - \Psi^{\dagger}_2 \Psi_1),    \nonumber    
\end{align}
with $\Psi_{l}$ ($l=1,2$) the annihilation operator for an electron on edge $l$ and where we neglected the time dependence for notational convenience.
Replacing these expressions into Eq.~(\ref{eq:noisestart})
allows to express the outgoing noise as~\cite{grenier_single_2011}
\begin{equation}
\mathcal{S}_{12}^{\, \mathrm{out}} =  \mathcal{R T} \left(\mathcal{S}_{11} + \mathcal{S}_{22} +Q \right).
\label{eq:S12}
\end{equation}
The last term encodes quantum interference contributions:
\begin{align}
Q = -e^2 \int dt dt' \;  &\Big[
    \langle \Psi_1(t) \Psi_1^{\dagger}(t') \rangle \langle \Psi_2^{\dagger}(t) \Psi_2(t') \rangle \nonumber \\
    & +  \langle \Psi_1^{\dagger}(t) \Psi_1(t') \rangle \langle \Psi_2(t) \Psi_2^{\dagger}(t') \rangle \Big], 
\end{align}
while the first two terms are the incoming auto-correlation noise associated respectively to $I_{1}$ and $I_{2}$.
Notice that the averages are evaluated with respect to the initial state $|\varphi\rangle$ and correspond to the first order electronic coherence functions defined in~\cite{grenier_single_2011}.

Calculations are performed in the simple case in which a single electron, with a given WP, is injected into each edge. This allows to obtain analytical expressions for the HOM dip, which corresponds to the cross-correlated noise when two WPs collide from opposite sides of the QPC.
The state corresponding to an electron injected with WP $\varphi_{1,2}(x)$ on edge $1, 2$ reads:
\begin{equation}
| \varphi_{1,2} \rangle = \int dx \; \varphi_{1,2}(x) \; \Psi^{\dagger}_{1,2}(x) \; | F \rangle
\label{eq:Psie}
\end{equation}
where $| F \rangle$ indicates the edge state Fermi sea at finite temperature $T$. 
For identical WPs $\varphi_{1,2}=\varphi(x)$, reaching the QPC with a delay $\delta t$, the noise normalized by twice the HBT noise (only one source emitting \cite{bocquillon_electron_2012a}) is:
\begin{equation}
\frac{\mathcal{S}_{HOM}(\delta t)}{2 \mathcal{S}_{HBT}} =
1 - \left| \frac{\int_0^\infty dk
  |\tilde{\varphi}(k)|^2 e^{-i k \delta t} (1-f_k)^2}
  {\int_0^\infty dk  |\tilde{\varphi}(k)|^2 (1-f_k)^2} \right|^2, 
\label{eq:HOMee1}
\end{equation}
where 
\begin{equation}
f_k = \frac{1}{1+e^{(k-k_F)/T}}
\end{equation}
is the Fermi distribution ($k_{F}$ the Fermi momentum),
$\tilde{\varphi}(k)$ is the WP in momentum space, and
\begin{equation}
 \mathcal{S}_{HBT}   =
-e^2 \mathcal{R} \mathcal{T}
\left(
\frac{\int_0^\infty dk
  |\tilde{\varphi}(k)|^2  (1-f_k)^2}
  {\int_0^\infty dk  |\tilde{\varphi}(k)|^2 (1-f_k)} \right)^2.
\label{eq:SHBT}
\end{equation}
Eq.~(\ref{eq:HOMee1}) shows immediately that $\mathcal{S}_{HOM}(0)/2 \mathcal{S}_{HBT}=0$, as a consequence of Fermi statistics. 
On the opposite, when $\delta t$  is much larger than the inverse width of $\tilde\varphi(k)$, $\mathcal{S}_{HOM}(\infty)$ is
the sum of the HBT noise of the two electrons taken independently, and $S_{HOM}/(2 S_{HBT})=1$. At low temperature, when the injected WP does not overlap with the Fermi sea, one has:
\begin{equation}
\frac{\mathcal{S}_{HOM}(\delta t)}{2 \mathcal{S}_{HBT}} =
  1- \left| \int d\tau \; \varphi(\tau) \, \varphi^*(\tau+\delta t)\right|^2,
  \label{eq:HOMeerealspace}
\end{equation}
which is similar to the case of optics \cite{hong_measurement_1987} as the profile of the HOM dip corresponds to the auto-convolution of the WP.
The functional forms of the HOM dip can be obtained analytically for various types of WPs.
The SES of Ref. \cite{feve_-demand_2007} is believed to generate Lorentzian WPs of the form:
\begin{equation}
\tilde\varphi(k) = \frac{\sqrt{\Gamma}}{\sqrt{2\pi}} \frac{1}{(k-k_0)+i \frac{\Gamma}{2}}
\label{emitted_wp}
\end{equation}
which corresponds to Breit-Wigner resonance associated with the emission by the discrete level of a quantum dot of width $\Gamma/2$ at energy $k_0$, with a half exponential (see Fig~\ref{fig1}) real space profile 
$\varphi(x) = \sqrt{\Gamma} e^{i k_0 x} e^{\frac{\Gamma}{2} x} \theta(-x)$
($\theta(x)$ is the Heaviside function).
At zero temperature, the noise corresponding to two such WPs, 
centered at the same energy $k_0$ but with different widths $\Gamma_{1,2}$
reads:
\begin{equation}
\frac{\mathcal{S}_{HOM}(\delta t)}{2 \mathcal{S}_{HBT}} = 1 -
 \frac{4 \Gamma_1 \Gamma_2}{(\Gamma_1 + \Gamma_2)^2}  
  \Big[ \theta(\delta t) e^{-\Gamma_1 \delta t} + \theta(-\delta t) e^{\Gamma_2 \delta t} \Big].
\label{eq:HOMeeExp}
\end{equation}

This HOM dip lacks mirror symmetry: its exponential behavior is characterized by the 
time constants $\Gamma^{-1}_1$ or $\Gamma^{-1}_2$ respectively depending on the sign of $\delta t$. 
The dip does not reach zero as for Eq. (\ref{eq:HOMee1}). The reduced contrast (less than unity) 
reflects the distinguishability of the injected electrons. 
This asymmetry is clearly not present for WPs generated by a Lorentzian voltage applied directly to the edge channels \cite{dubois_integer_2013,dubois_minimal_2013,jullien_quantum_2014} or when considering the adiabatic limit for the emission of a SES \cite{olkhovskaya_shot_2008,splettstoesser_two_2009,haack_coherence_2011}.
 
We compare our formulas with the numerical results of 
a Floquet calculation \cite{moskalets_floquet_2002,moskalets_noise_2013,moskalets_single_2016} properly modeling the emission process from a realistic periodic source~\cite{moskalets_quantized_2008,parmentier_current_2012}.
This emitter consists of the mesoscopic capacitor~\cite{feve_-demand_2007,mahe_current_2010,bocquillon_electron_2012a}: a quantum dot with discrete levels connected through a QPC to the edge state which is driven by a gate applying a periodic square drive $V(t)$.  
The highest occupied state is first positioned above the Fermi level, causing the tunneling of a dot electron to the edge; this (now) empty level
is next placed below the Fermi sea, causing the emission of a hole. 
Periodic square voltage with an amplitude identical to the dot level spacing $\Delta$ yield optimal emission. 

  \begin{figure}[tbp]
\includegraphics[width=9.cm]{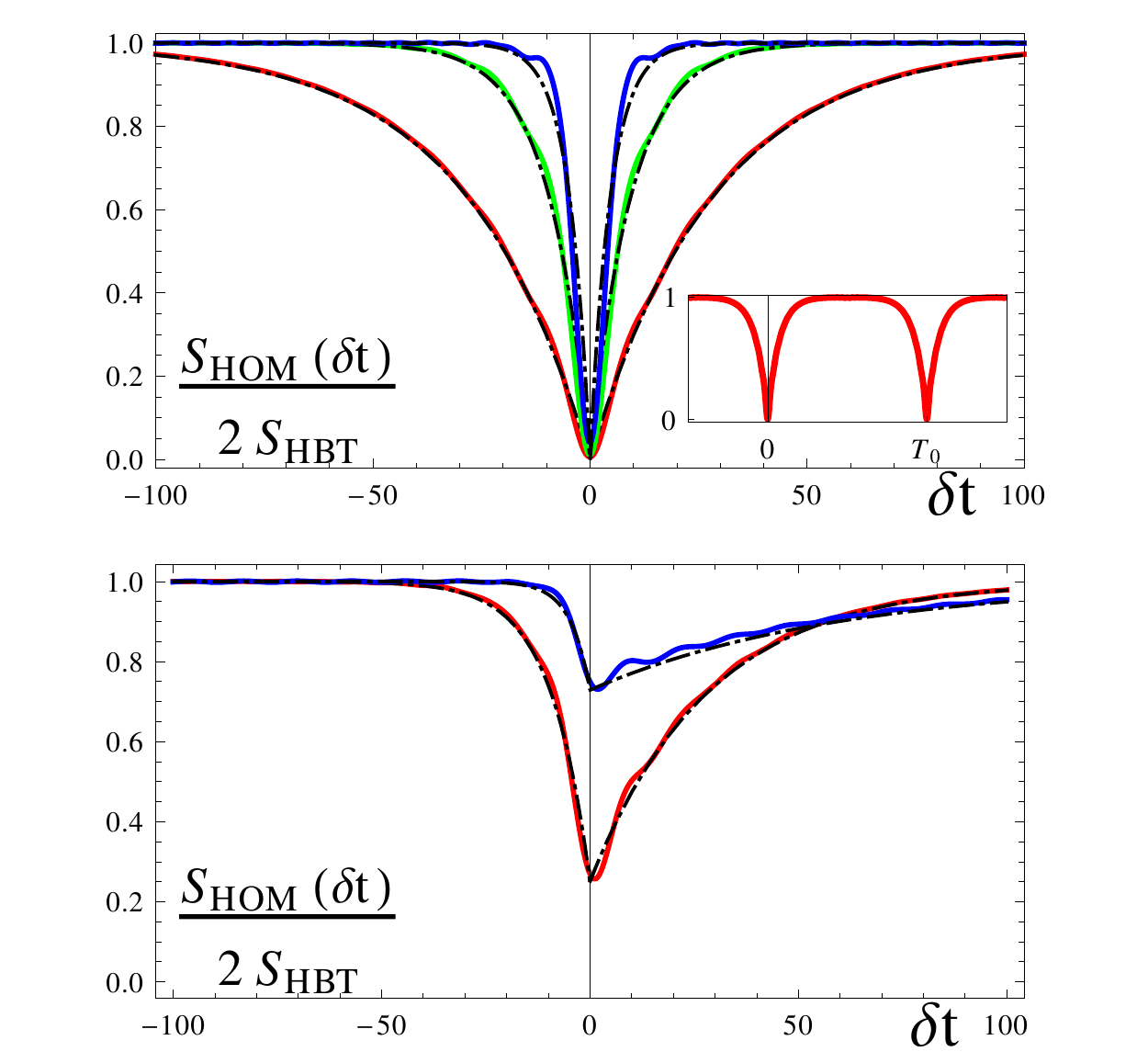}
\caption{(Color online) Normalized HOM noise as a function of the delay in the electron (hole) emission $\delta t$. Full curves represent the result obtained within the Floquet scattering matrix formalism in the optimal emission regime of the SES. Dotted-dashed curves are the analytical predictions of Eq.~(\ref{eq:HOMeeExp}) for exponential WPs. 
Upper panel: symmetric profile, with emitter transparencies $D=0.2$ (red curve), $0.5$ (green curve) and $0.8$ (blue curve).
Inset: dips for $D=0.2$ on two periods of the applied voltage.
 Lower panel: asymmetric profile, with transparencies
 $D_1=0.2$, $D_2=0.5$ (bottom red curve) and $D_1=0.1$, $D_2=0.8$ (top blue curve). Other parameters are: $T_0 = 400$ (in units of $\hbar/\Delta$) the period of the applied voltage, and $T=0.01 \Delta$ the temperature. Picture taken from \cite{jonckheere_electron_2012}.}
\label{fig2}
\end{figure} 

The Floquet approach allows to evaluate numerically current and noise \cite{parmentier_current_2012}.  
Fig.~\ref{fig2} shows the comparison between the numerical results for the HOM dip with
the analytical formula in Eq.~(\ref{eq:HOMeeExp}). 
The upper panel shows a symmetric profile, due to the fact that the two SES are identical.
The dips (with minimum value $0$ at zero delay) are broader for lower emitter transparencies. This is a consequence of the fact that electrons take
a longer time to exit the dot leading to a broader WP.
In the optimal regime, the electron emission time corresponds to \cite{parmentier_current_2012,nigg_quantum_2008}. 
\begin{equation}
\tau = \frac{2\pi}{\Delta} \left(\frac{1}{D}- \frac{1}{2}\right).
\end{equation}
This value (with $\Gamma=\tau^{-1}$) is chosen to plot the analytical
predictions from Eq.~(\ref{eq:HOMeeExp}) (dotted-dashed curves).
The agreement is excellent, especially at low enough transparency, where
the single electron emission is properly achieved~\cite{mahe_current_2010,grenier_single_2011}.
For what it concerns the asymmetric profile (lower panel), the contrast is smaller than 1. Also in this case both the numerical and the analytical approaches agree very well.
 
The HOM interferometry with fermions is thus characterized by a dip in the zero-frequency current cross-correlations in collisions between two electrons, whose shape depends on the injected WPs. This same setup also offers the interesting possibility (not shown) to achieve electron hole collisions, which produce an HOM peak at large enough temperature \cite{jonckheere_electron_2012}.

\section{Electronic quantum optics with 2D topological insulators.}
Such materials exhibit the QSH effect \cite{hasan_colloquium_2010,qi_topological_2011}. The first experimental observations of this peculiar state of matter have been carried out in CdTe/HgTe \cite{bernevig_quantum_2006,konig_quantum_2007} and InAs/GaSb
\cite{liu_experimental_2009,knez_evidence_2011,du_robust_2015} quantum wells. QSH effect manifests itself through a gapped bulk and metallic edge states \cite{wu_helical_2006}
where electrons with opposite spin propagate in opposite directions along the edges as a consequence of spin-orbit interaction. The topologically protected helical edge states of QSH effect, with their spin-momentum locking properties, suggests that they could be studied in a EQO context \cite{ferraro_electronic_2014} especially given the recent proposals for an electron source and beam-splitter. Indeed, the characterization of the SES has already been provided \cite{hofer_emission_2013,inhofer_proposal_2013}. This pair electron source (PES) are predicted to trigger the injection into the helical edge states of a pair of electrons (holes) with opposite spin per period. However, the experimental realization of a QPC in the QSH regime still represents a challenge due to the same Klein mechanism which prevents confinement of massless Dirac fermions in graphene \cite{katsnelson_chiral_2006}. Still, new generation heterostructures \cite{knez_evidence_2011,du_robust_2015} give reasonable hopes of possible applications to EQO.
\begin{figure}[ht]
\centering
\includegraphics[scale=0.35]{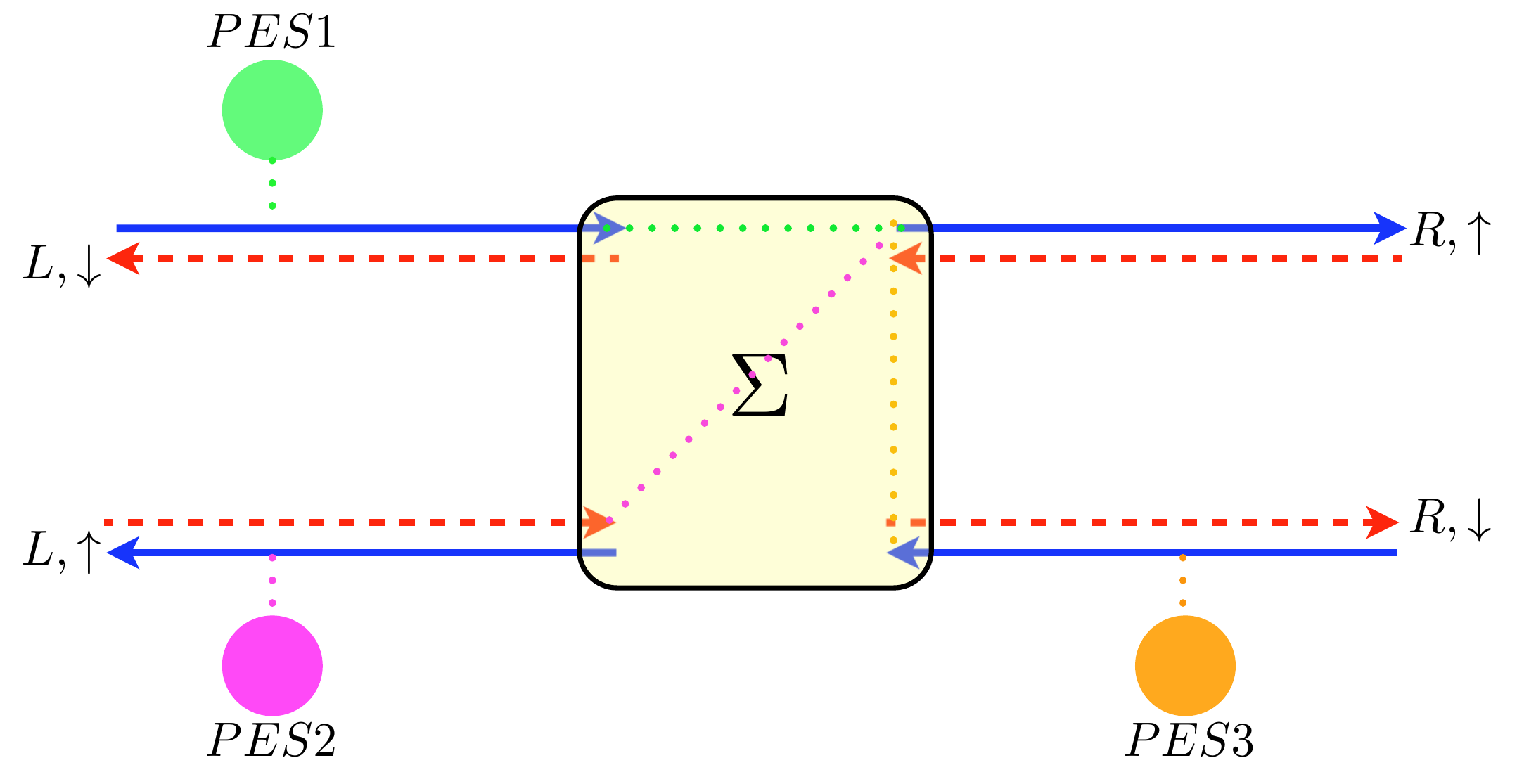}
\caption{(Color on line) Schematic view of a QSH bar with spin up (full arrows) and spin down (dashed arrows) electrons. Dotted lines indicate the scattering processes, compatible with TRS, which affect the current in the $R,\uparrow$ outgoing channel (see main text). Pair-electrons sources (PES) are represented with colored circles. Picture taken from \cite{ferraro_electronic_2014}.}
\label{fig3}
\end{figure}

\subsection{Model.}\label{model_topo}
The Hamiltonian of the system, in the presence of a QPC, is given by \cite{inhofer_proposal_2013,dolcini_full_2011,citro_electrically_2011,romeo_electrical_2012}
  \begin{equation}
 \mathcal{H}= \mathcal{H}_{0}+ \mathcal{H}_{sp}+\mathcal{H}_{sf}
 \end{equation}
with
\begin{equation}
\mathcal{H}_{0}= -i \hbar \sum_{\alpha=R, L}  \sum_{\sigma= \uparrow, \downarrow} \int^{+\infty}_{-\infty} dx \xi_{\alpha} : \Psi^{\dagger}_{\alpha, \sigma}(x) \partial_{x} \Psi_{\alpha, \sigma}(x): 
\end{equation}
the free Dirac Hamiltonian of the one-dimensional helical edge channels. Here $\Psi_{\alpha, \sigma}(x)$ extends the previous definition for the electronic annihilation operator by taking into account the chirality ($\alpha=R, L$) and spin ($\sigma=\uparrow, \downarrow$) degrees of freedom. Moreover, $\xi_{R/L}=\pm 1$ represents the chirality index and $:...:$ indicates the conventional normal ordering. Assuming a local QPC, one obtains two additional contributions: 
\begin{equation}
\mathcal{H}_{sp}= 2 \hbar  \sum_{\sigma= \uparrow, \downarrow} \gamma_{sp} \Psi^{\dagger}_{R, \sigma}(0)\Psi_{L, \sigma}(0)+h. c. 
\label{Hsp}
\end{equation}
the spin-preserving and
\begin{equation}
\mathcal{H}_{sf}= 2 \hbar  \sum_{\alpha= R, L} \xi_{\alpha} \gamma_{sf}\Psi^{\dagger}_{\alpha, \uparrow}(0) \Psi_{\alpha, \downarrow}(0)+h.c.
\label{Hsf}
\end{equation}
the spin-flipping tunneling Hamiltonian.

TRS of the total Hamiltonian $\mathcal{H}$ is guaranteed as long as $\gamma_{sp}$ and $\gamma_{sf}$ are real numbers \cite{dolcini_full_2011,ferraro_spin_2013} ($\gamma_{sp} > \gamma_{sf}$ is generally assumed). 
The evolution of the fermionic field operators is specified by the Heisenberg equation of motion in Dirac form. This allows to specify the incoming/outgoing scattering states/operators, which are related by a $4\times 4$ scattering matrix describing the QPC \cite{dolcini_full_2011,delplace_magnetic_2012}, whose structure reflects the TRS. The scattering matrix elements are then parametrized by 
$\lambda_{pb}$,
$\lambda_{ff}$,
$\lambda_{pf}$ \cite{dolcini_full_2011,delplace_magnetic_2012}, namely the amplitude probabilities of spin-preserving backscattering, spin-flipping forward scattering and spin-preserving forward scattering processes (respectively orange, magenta and green dotted lines in Fig.~\ref{fig3}) \cite{inhofer_proposal_2013}. They naturally satisfy the constraint
\begin{equation}
|\lambda_{ff}|^{2}+\lambda_{pf}^{2}+|\lambda_{pb}|^{2}=1
\end{equation}
imposed by charge conservation.

\subsection{Auto-correlated noise.}\label{Current_Noise}
We want to investigate now the auto-correlated outgoing noise. We will focus for simplicity on the ($R, \uparrow$) channel, the expressions for the other possible cases can be derived proceeding in the same way. It reads
\begin{equation}
\mathcal{S}^{\, \mathrm{out}}_{R \uparrow, R\uparrow}= \int dt dt' \langle I^{\, \mathrm{out}}_{R\uparrow}(t) I^{\, \mathrm{out}}_{R\uparrow}(t')\rangle_{\rho, c}
\end{equation}
where the notation indicates that the connected correlator is calculated over a density matrix of the form $\rho=|\varphi \rangle \langle \varphi|$.
In terms of the incoming signals, it can be written as
\begin{eqnarray}
\mathcal{S}^{\, \mathrm{out}}_{R\uparrow, R\uparrow}&=& |\lambda_{ff}|^{4} \mathcal{S}_{R\downarrow, R\downarrow} +\lambda^{4}_{pf}\mathcal{S}_{R\uparrow, R\uparrow}\nonumber\\
&+&|\lambda_{pd}|^{4}
\mathcal{S}_{L \uparrow, L\uparrow}+Q.
\label{S_out}
\end{eqnarray}
Notice that the interesting physics is encoded in the last term of Eq.~(\ref{S_out}) which extends what we already evaluated in the IQH case, while the others are the auto-correlations of the incoming currents, which vanish when taking the average over one period and computing their zero frequency Fourier component \cite{mahe_current_2010,bocquillon_electron_2012a}. Its explicit form here is
\begin{eqnarray}
Q&=& \left[ (\mathcal{A}+\mathcal{B}+\mathcal{C}) Q^{(FS)}+(\mathcal{A}+\mathcal{B}) Q^{(HBT)}_{R\downarrow}  \right.  \nonumber\\
&~&+ (\mathcal{A}+\mathcal{C}) Q^{(HBT)}_{R\uparrow}+ (\mathcal{B}+\mathcal{C}) Q^{(HBT)}_{L\uparrow}\nonumber\\
&~& \left. + \mathcal{A} Q^{(HOM)}_{R\downarrow, R \uparrow}+\mathcal{B} Q^{(HOM)}_{R\downarrow, L \uparrow}+ \mathcal{C} Q^{(HOM)}_{R\uparrow, L \uparrow}\right]
\label{Q}
\end{eqnarray}
where 
$\mathcal{A}= |\lambda_{ff}|^{2} \lambda^{2}_{pf}$, $\mathcal{B}= |\lambda_{ff}|^{2} |\lambda_{pb}|^{2}$,
$\mathcal{C}=\lambda^{2}_{pf} |\lambda_{pb}|^{2}$ (these can be tuned by modifying the QPC parameters \cite{romeo_electrical_2012,krueckl_switching_2011}). The Fermi sea \cite{blanter_shot_2000,martin_noise_2005}, the HBT \cite{bocquillon_electron_2012a}, and the HOM \cite{jonckheere_electron_2012,bocquillon_coherence_2013} noise contributions read respectively:
\begin{align}
Q^{(FS)}= &\frac{e^{2}}{\pi} \int d \bar{t} d \xi f_{\xi} \left(1- f_{\xi} \right)\\
Q^{(HBT)}_{a}= &\frac{e^{2}}{2\pi} \int d \bar{t} d \xi \Delta \mathcal{W}^{(e)}_{a}(\bar{t}, \xi) \left(1- 2f_{\xi} \right)\\
Q^{(HOM)}_{a, b}= &-\frac{e^{2}}{\pi} \int d \bar{t} d \xi \Delta \mathcal{W}^{(e)}_{a}(\bar{t}, \xi) \Delta \mathcal{W}_{b}(\bar{t}+\delta, \xi)
\label{noise_contrib}
\end{align}
with $a$ and $b$ the channels of injection and
\begin{equation}
\Delta \mathcal{W}^{(e)}_{a}(\bar{t}, \xi)=\int d \tau e^{i \xi \tau} \Delta\mathcal{G}^{(e)}_{a}\left(\bar{t}+\frac{\tau}{2}, \bar{t}-\frac{\tau}{2} \right) 
\end{equation}
the Wigner function \cite{ferraro_wigner_2013} obtained as a partial Fourier transform of the excess first order coherence \cite{grenier_single_2011}
\begin{equation}
\Delta \mathcal{G}^{(e)}_{a}(t, t')= \langle \Psi^{\dagger}_{a}(t') \Psi_{a}(t) \rangle_{\rho}-\langle F | \Psi^{\dagger}_{a}(t') \Psi_{a}(t) |F\rangle .
\end{equation}

In the absence of spin-flip ($\mathcal{A}=\mathcal{B}=0$), $\mathcal{C}$ becomes the product of the transmission and reflection probability of the QPC and one recovers what is observed in the IQH case. On a more general ground, when one of the scattering amplitudes is zero only one of the $\mathcal{A}$, $\mathcal{B}$, $\mathcal{C}$ parameters survives and we recover an equivalent result to that of the IQH situation.

In experiments, one subtracts the Fermi sea contributions to define the excess noise: 
\begin{equation}
\Delta Q= Q- (\mathcal{A}+\mathcal{B}+\mathcal{C}) Q^{(FS)}.
\end{equation}
When considering the emission of a pair of identical WPs in the form of Eq. (\ref{emitted_wp}) from a PES, the HBT contributions reduce to $Q^{(HBT)}_{a}\approx e^{2}$ while the HOM contributions \cite{jonckheere_electron_2012} read
$Q^{(HOM)}_{a, b}(\delta t) \approx -2 e^{2} \exp{\left(-\Gamma |\delta t|\right)}$. Notice that, for the sake of simplicity, we have neglected the overlap of the injected electron WP with the Fermi distribution of the other channels (emission high above the Fermi sea and at very low temperature \cite{bocquillon_electron_2012a}).

\begin{figure*}[tbp]
\centering
\includegraphics[scale=0.55]{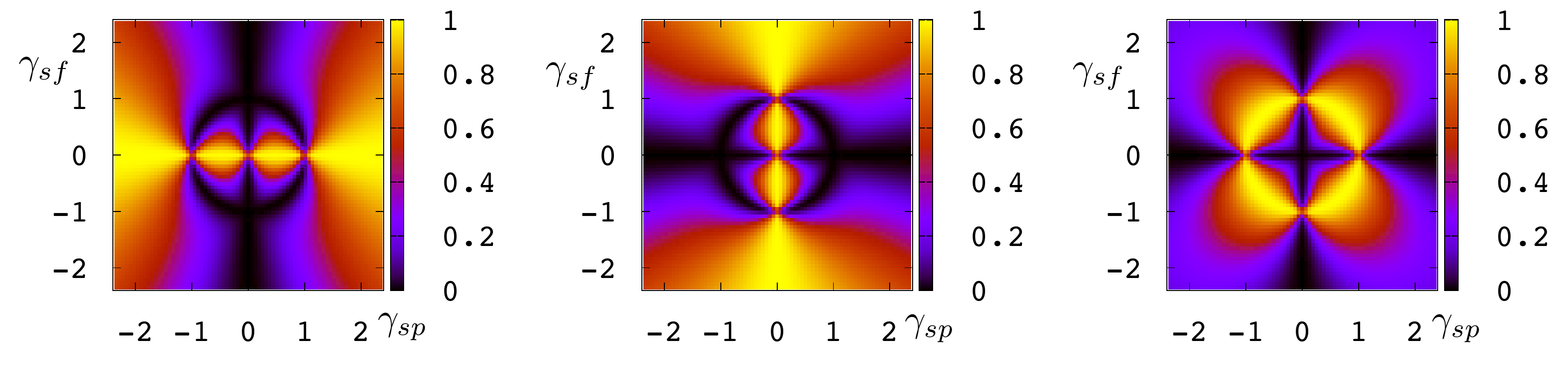}
\caption{(Color on-line) Density plot of $\mathcal{I}$ (left), $\mathcal{J}$ (middle) and $\mathcal{K}$ (right) as a function of $\gamma_{sp}$ and $\gamma_{sf}$. Picture taken from \cite{ferraro_electronic_2014}.}
\label{fig4}
\end{figure*}

\subsubsection{Two-electron collision.}
\begin{figure}[ht]
\centering
\includegraphics[scale=0.35]{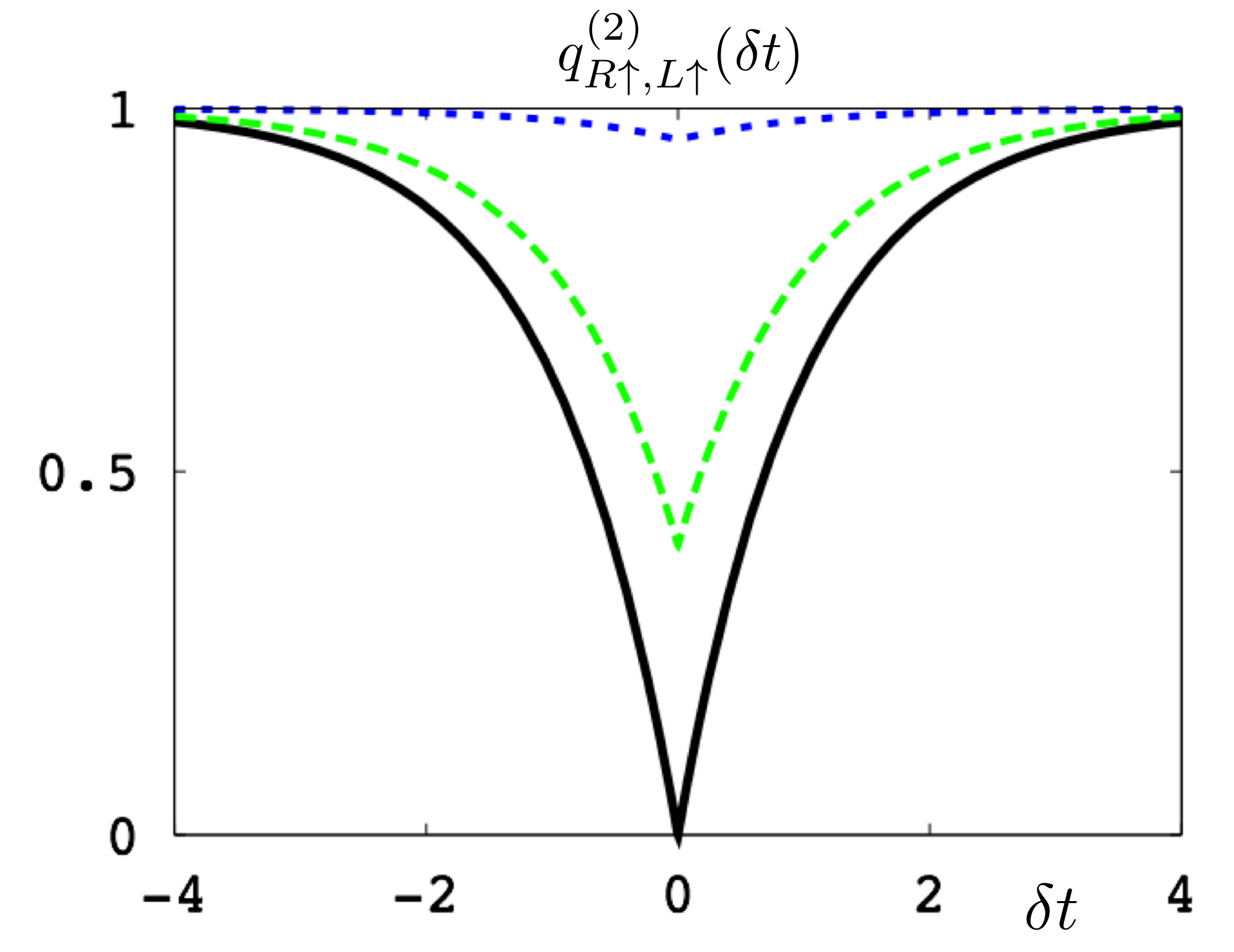}
\caption{(Color on-line) Behavior of $q^{(2)}_{R\uparrow, L\uparrow}(\delta t)$ as a function of $\delta t$ (in units of $\Gamma^{-1}$) for different values of spin-flipping and spin-preserving amplitudes: $\gamma_{sp}=2, \gamma_{sf}=0$ (full black curve), $\gamma_{sp}=2, \gamma_{sf}=1.5$ (dashed green curve) and $\gamma_{sp}=1, \gamma_{sf}=0.3$ (dotted blue curve). Picture taken from \cite{ferraro_electronic_2014}.}
\label{fig5}
\end{figure}

We consider the injection of electrons into the $(R,\uparrow)$ and the $(L,\uparrow)$ incoming channels. Here, only $PES1$ and $PES3$ (respectively green and orange circles in Fig.~\ref{fig3}) are ``on''. This process is therefore the QSH equivalent of the IQH case (equal spin injection). The more relevant physical quantity to look at is the ratio between the HOM noise (two sources emitting together with finite delay $\delta t$) and the sum of the HBT noises associated with the same sources: 
\begin{equation}
q^{(2)}_{R\uparrow, L\uparrow}(\delta t)
\approx 1- \mathcal{I}e^{-\Gamma |\delta t|}
\label{q2_upup}
\end{equation}  
where $\mathcal{I}= 2\mathcal{C}/(\mathcal{A}+\mathcal{B}+2\mathcal{C})$
is the visibility (see Fig.~\ref{fig4}). 
Eq.~(\ref{q2_upup}) predicts a dip in the noise for electrons reaching the QPC with a delay such as $\Gamma|\delta t| < 1$. Moreover, the exponential form of the dip is reminiscent of the WP profile. As in the IQH case discussed above, this \emph{Pauli dip} is due to the fermionic statistics of the electrons \cite{bocquillon_coherence_2013}, while the reduced visibility (compared to \cite{jonckheere_electron_2012}) is due to the presence of additional channels coupled at the QPC \cite{rizzo_transport_2013}. Indeed, more outgoing channels lead to more partitioning at the QPC and a consequent enhancement of the HBT noise contribution.
This effect can be very small or conversely quite important depending on the visibility $\mathcal{I}$ (and consequently of the intensity of $\gamma_{sp}$ and $\gamma_{sf}$) as shown in Fig.~\ref{fig5}. Notice that, for $\gamma_{sf}=0$ (absence of spin-flipping) one has $\mathcal{I}=1$ and we recover the result of the IQH.
It is worth mentioning that the suppression of the visibility discussed here has a different physical origin with respect to the one observed in the IQH effect at filling factor $\nu=2$ \cite{bocquillon_coherence_2013,wahl_interactions_2014}. Indeed, even if in this case several scattering channels are present, the QPC can be easily experimentally  tuned in a region where this effect is absent (partial transmission of the outer channel and total reflection of the others). Moreover, additional checks allows to unequivocally identify the inter-channel interaction as the dominant cause for the loss of contrast in the IQH case \cite{marguerite_decoherence_2016}.

Due to spin flip processes occurring at the QPC, opposite spin electron interferometry is also possible for this kind of device. For electrons of the same chirality we have: 
\begin{equation}
q^{(2)}_{R\downarrow, R\uparrow}(\delta t)\approx 1- \mathcal{J} e^{-\Gamma |\delta t|}
\label{q2_RR}
\end{equation}
when $PES1$ and $PES2$ are ``on'' (green and magenta circles in Fig.~\ref{fig3}). Notice that we have defined a new visibility factor
$\mathcal{J}= 2\mathcal{A}/(2\mathcal{A}+\mathcal{B}+\mathcal{C})$ (see Fig.~\ref{fig4}).
It is clear that the interference process between $(R,\uparrow)$ and $(L,\uparrow)$ electrons can be mapped into the one involving $(R,\uparrow)$ and $(R,\downarrow)$ electrons by exchanging the spin-preserving and spin-flipping contributions. Notice that at $\gamma_{sp}=0$ ($\mathcal{B}=\mathcal{C}=0$) we achieve the maximum visibility ($\mathcal{J}=1$). 

\begin{figure}[ht]
\centering
\includegraphics[scale=0.35]{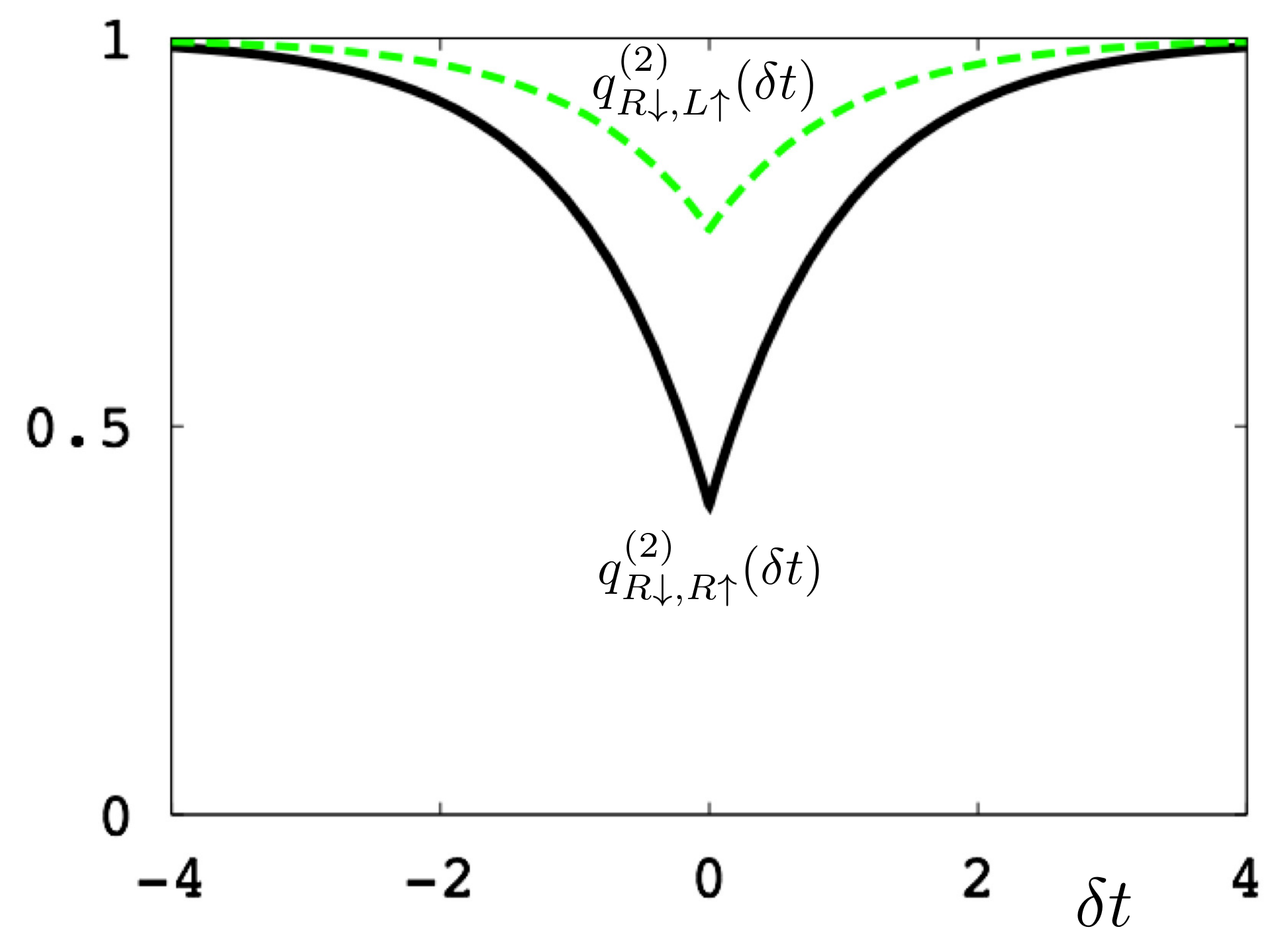}
\caption{(Color on-line) Behavior of $q^{(2)}_{R\downarrow, R\uparrow}(\delta t)$ (full black) and $q^{(2)}_{R\downarrow, L\uparrow}(\delta t)$ (dashed green) as a function of the $\delta t$ (in units of $\Gamma^{-1}$). Parameters are $\gamma_{sp}=\gamma_{sf}=2$. Picture taken from \cite{ferraro_electronic_2014}.}
\label{fig6}
\end{figure}

The present setup offers also the novel possibility to realize the interference of electrons with opposite spin and opposite chirality:
\begin{equation}
q^{(2)}_{R\downarrow, L\uparrow}(\delta t)\approx 1-\mathcal{K} e^{-\Gamma |\delta t|}
\label{q2_downup}
\end{equation}
with $PES2$ and $PES3$ turned ``on'' (magenta and orange circles in Fig.~\ref{fig3}) and 
$\mathcal{K}= 2\mathcal{B}/(\mathcal{A}+2\mathcal{B}+\mathcal{C})$. Maximum visibility is reached when $\mathcal{A}=\mathcal{C}=0$, namely in a circle of radius 1 in the ($\gamma_{sp},\gamma_{sf}$) plane (see Fig.~\ref{fig4}).

In the three different two-electron collision configurations (Eqs.~(\ref{q2_upup}), (\ref{q2_RR}) and (\ref{q2_downup})) the maximal visibility occurs when one of the scattering amplitudes (respectively $\lambda_{ff}$, $\lambda_{pb}$ or $\lambda_{pf}$) is zero. Indeed, in this case, only two outgoing channels are available for the two electrons and we recover a zero noise as in the IQH case. 

However, the noise suppression observed for collision of electrons with opposite spin is by far not trivial and is due to the constraints imposed by TRS and charge conservation in the QSH system (see Ref.~\cite{edge_$z_2$_2013} for a similar discussion in the case of a continuous current). This phenomenon, known as $\mathcal{Z}_{2}$ dip, has also been discussed in Ref.~\cite{inhofer_proposal_2013} for a specific range of parameters ($\gamma_{sp}=0$, $\gamma_{sf}\neq0$), a particular case of the more general analysis reported here. 

\subsubsection{Three-electron collision.}
Unique to the QSH effect is the possibility  to translate three-photon HOM experiments \cite{campos_three_2000} to EQO. Here, the three PES of the setup in Fig.~\ref{fig3} are operating. The delays in the electron emission are respectively $\delta t_{1}$ between $(R,\downarrow)$ and $(R,\uparrow)$; $\delta t_{2}$ between $(R,\downarrow)$ and $(L,\uparrow)$. Consequently, $(\delta t_{2}-\delta t_{1})$ represents the delay between $(R,\uparrow)$ and $(L,\uparrow)$. As previously we obtain a normalized noise:
\begin{eqnarray}
&&q^{(3)}(\delta t_{1}, \delta t_{2}) 
\approx  1- \frac{\mathcal{A}}{\mathcal{A}+\mathcal{B}+\mathcal{C}} e^{-\Gamma |\delta t_{1}|}\nonumber\\
&-&\frac{\mathcal{B}}{\mathcal{A}+\mathcal{B}+\mathcal{C}} e^{-\Gamma |\delta t_{2}|}-\frac{\mathcal{C}}{\mathcal{A}+\mathcal{B}+\mathcal{C}} e^{-\Gamma |\delta t_{2}-\delta t_{1}|}.
\end{eqnarray}  
Remarkably enough, a perfect synchronization between the PES leads to $q^{(3)}(\delta t_{1}=0, \delta t_{2}=0)=0$
\emph{for an arbitrary QPC}. This total noise suppression is a consequence of the interplay between the fermionic statistics and the TRS in the QSH systems. This can be understood by considering the input state $a_{R\uparrow}^\dagger a_{R\downarrow}^\dagger a_{L\uparrow}^\dagger | F \rangle$ (creation of three electrons simultaneously, on in each input channel). 
In terms of the scattering matrix, one can rewrite this as:
\begin{eqnarray}
a_{R\uparrow}^\dagger a_{R\downarrow}^\dagger a_{L\uparrow}^\dagger  | F \rangle &=& \left( \lambda_{pb}^* b_{L\uparrow}^\dagger + \lambda_{pf} b_{R\uparrow}^\dagger + \lambda_{ff}^* b_{R\downarrow}^\dagger \right) \nonumber \\
&~& \times \left( \lambda_{pb}^* b_{L\downarrow}^\dagger + \lambda_{ff}^* b_{R\uparrow}^\dagger + \lambda_{pf} b_{R\downarrow}^\dagger \right) \nonumber \\
&~& \times \left( \lambda_{pf} b_{L\uparrow}^\dagger + \lambda_{ff} b_{L\downarrow}^\dagger + \lambda_{pb}^* b_{R\uparrow}^\dagger \right) | F \rangle.
\nonumber\\
\end{eqnarray}
TRS guarantees that each incoming operator $a_{\alpha\sigma}^\dagger$ can be expressed as a linear combination of only 3 out of the 4 outgoing operators $b_{\alpha\sigma}^\dagger$. Therefore, exploiting the unitarity of the scattering matrix and the Pauli principle one obtains
\begin{eqnarray}
a_{R\uparrow}^\dagger a_{R\downarrow}^\dagger a_{L\uparrow}^\dagger | F \rangle =& \left( \lambda_{pb} b_{L\uparrow}^\dagger  b_{L\downarrow}^\dagger   
+ \lambda_{pf} b_{R\downarrow}^\dagger  b_{L\uparrow}^\dagger   
\right. \nonumber \\
&  \left. + \lambda_{ff} b_{R\downarrow}^\dagger  b_{L\downarrow}^\dagger  \right) 
b_{R\uparrow}^\dagger| F \rangle.
\end{eqnarray}
As it is easy to note, this leads to the superposition of three outgoing states, each involving the creation of an electron in the $(R,\uparrow)$ outgoing channel. Consequently this channel is always populated (no current fluctuations) independently of the final outcome of the scattering process. The partition noise in this channel vanishes, which constitutes a direct consequence of the interplay between TRS and Fermi statistics.

\begin{figure}[h]
\centering
\includegraphics[scale=0.50]{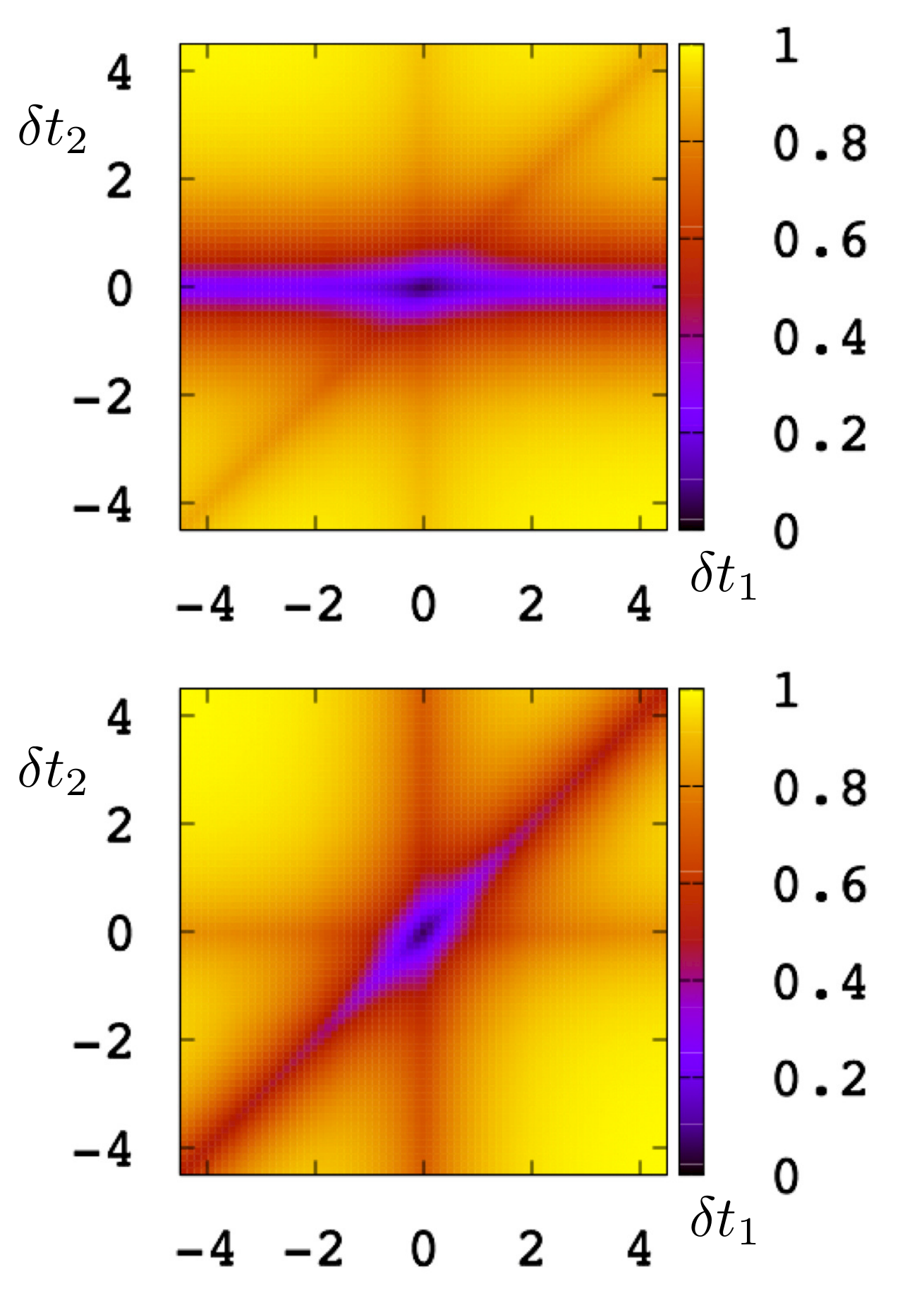}
\caption{(Color on-line) Density plot of $q^{(3)}(\delta t_{1}, \delta t_{2})$ as a function of $\delta t_{1}$ and $\delta t_{2}$ (in units of $\Gamma^{-1}$) for $\gamma_{sp}=1, \gamma_{sf}=0.8$ (top) and $\gamma_{sp}=2, \gamma_{sf}=1.5$ (bottom). Picture taken from \cite{ferraro_electronic_2014}.}
\label{fig7}
\end{figure}

In the absence of synchronized injections the phenomenology is even richer and crucially depends on the QPC parameters ($\gamma_{sp}$ and $\gamma_{sf}$). Situations may occur where the noise suppression is dominated by the ($R,\downarrow | L,\uparrow$) interference channel (see top panel of Fig.~\ref{fig7}), or oppositely equal spin injection ($R,\uparrow | L,\uparrow$) can be favored  (see bottom panel of Fig.~\ref{fig7}). According to this, the present setup allows to investigate different interference configurations, by changing the function of the QPC. Alternatively, HOM interferometry can be seen as a way to quantify the characteristics of the QPC, and in particular to assess the relative weight between the spin-preserving and the spin-flipping tunneling amplitudes.

To summarize, 2D topological insulators exhibiting the QSH effect, are characterized by a very rich physics related to the peculiar connection between spin and momentum of the electrons propagating along the edges. Equal spin, as well as opposite spin interference are allowed here as a consequence of TRS. Moreover, three-electron injection/interference phenomena similar to the ones observed for photons in the conventional quantum optics can be investigated in this setup, differently from what happens in the IQH case \cite{ferraro_electronic_2014}. 

\section{Non-local interference and Hong-Ou-Mandel collisions of individual Bogoliubov quasiparticles.}
\begin{figure}[h]
\centering
\includegraphics[scale=0.35]{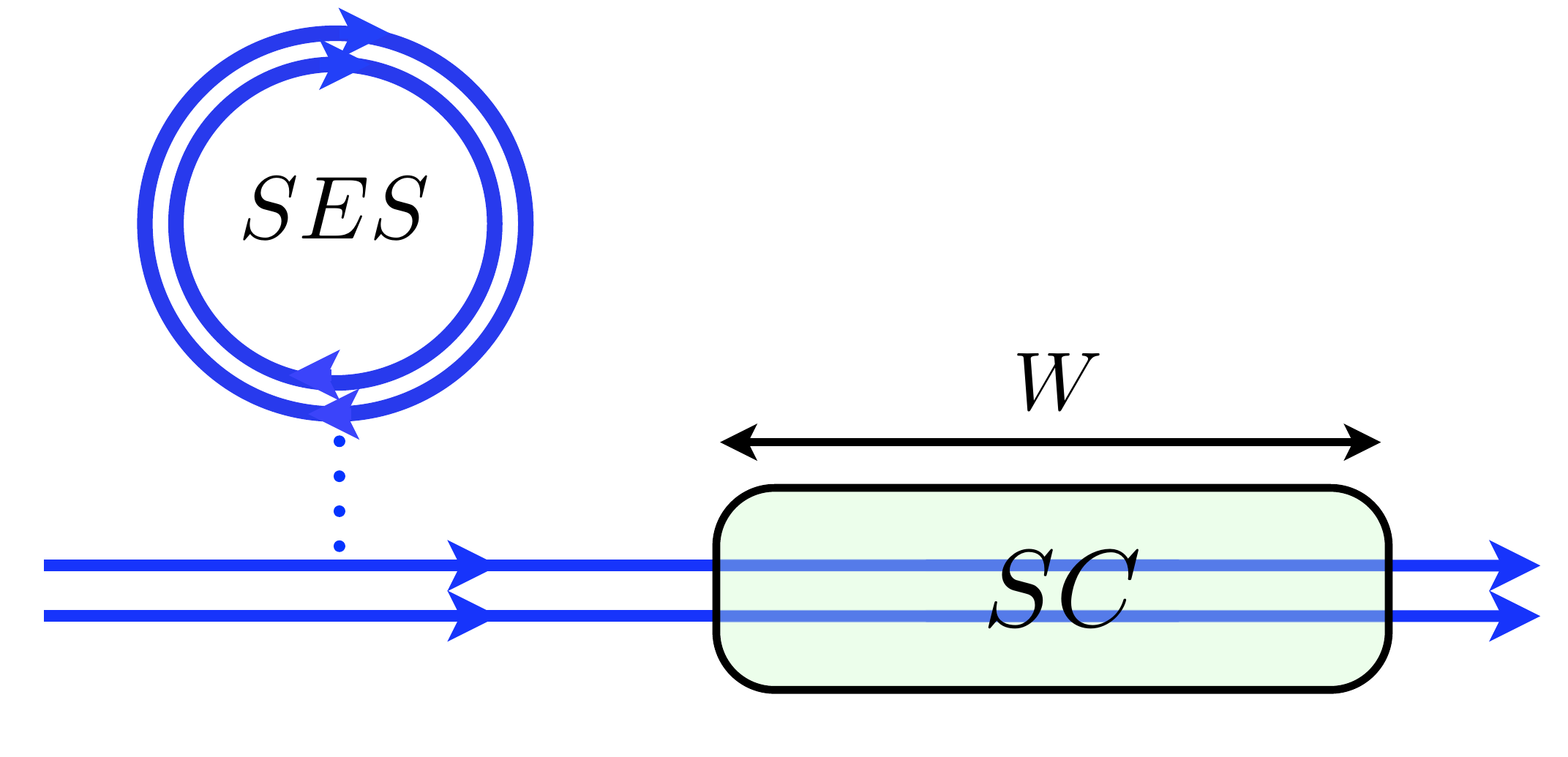}
\caption{(Color online) Schematic view of a single electron source (SES) injecting electron and hole WPs into a IQH edge state at filling factor $\nu=2$ coupled with a SC contact of length $W$. Picture taken from \cite{ferraro_nonlocal_2015}.}
\label{fig8}
\end{figure}

EQO scenarios are now revisited using a SES in the IQH regime put in proximity with a superconductor (SC) (see Fig. \ref{fig8}). The electrons which are injected can perform several Andreev reflections \cite{andreev_thermal_1964} and can thus be converted partially or totally into holes. Collisions between two Bogoliubov quasiparticles at the location of a QPC can be obtained in the framework of HOM interferometry. Ref. \cite{beenakker_annihilation_2014}, considered this setup first, with electron ``beams'' (DC voltage imposed between the two opposite edges) impinging on the QPC rather than single quasiparticle excitations. It has been argued \cite{chamon_quantizing_2010,beenakker_annihilation_2014} that since Bogoliubov quasiparticle creation operators are related by a unitary transformation to their annihilation counterpart, these excitations qualify as Majorana fermionic excitations. This constitutes an alternative proposal for Majorana fermions \cite{majorana_teoria_2008} compared to their topological superconductor counterparts \cite{akhmerov_electrically_2009,alice_new_2012} which give rise to a zero bias anomaly in tunneling experiments \cite{mourik_signatures_2012,das_zero_2012}. 
The DC proposal \cite{beenakker_annihilation_2014} failed to address the single shot creation and collision of two Bogoliubov quasiparticles, leading in principle to the annihilation of single Majorana excitations at the QPC. 
Moreover, this proposal requires the measurement of high frequency noise \cite{lesovik_detection_1997} in a normal metal/SC device.  More than a decade ago the finite frequency noise of a normal metal/SC junction was computed \cite{torres_effective_2001}, but it has so far eluded experimental observation. Single particle/quasiparticle injection, presents the great advantage that only zero frequency noise needs to be measured \cite{mahe_current_2010}. Here, the injection process is first characterized in terms of current and noise, exhibiting the non-conservation of the charge and the conservation of the excitation number. Next, EQO interferometric configurations with a QPC are studied, and it is shown that the averaged current (first order coherence \cite{grenier_electron_2011}) is independent of the SC phase difference, while oscillations dependent on the SC phase difference appear in the noise (second order coherence \cite{moskalets_two_2014,thibierge_two_2016}), which constitutes a clear signature of non local phenomena \cite{ferraro_nonlocal_2015} together with the annihilation of Majorana excitations at the QPC.

\subsection{Source of Bogoliubov quasiparticles.}\label{model_bogo}

Two edge channels at $\nu=2$ (Zeeman splitting and inter-channel interaction are ignored \cite{giazotto_andreev_2005}) are coupled to a SES and to a SC contact of length $W$ (see Fig. \ref{fig8}). The SES injects into the channels an electron (a hole) with well defined  WPs. A spin-singlet coupling between the Hall channels and the SC contact can for instance be realized in graphene \cite{popinciuc_zero_2012,rickhaus_quantum_2012}. 

The action of the SC on incoming electrons with energy below the induced SC gap $\Delta_{s}$ is described in terms of an energy dependent $4\times 4$ transfer matrix $\mathcal{M}$, constrained by unitarity and particle-hole symmetry \cite{beenakker_annihilation_2014,ostaay_spin_2011}:

\begin{equation}
\mathcal{M}(\xi)= \left(\tau_{x}\otimes I\right) \mathcal{M}^{*}(-\xi) \left(\tau_{x}\otimes I \right). 
\label{particle_hole}
\end{equation}
In the above expression  $I$ is the identity in spin space, while from now on we indicate with $\tau_{i}$ ($i=x, y, z$) the Pauli matrices acting on the electron-hole space and with $\sigma_{i}$ ($i=x, y, z$) the ones related to the spin degree of freedom. According to this, the transfer matrix $\mathcal{M}$ is applied to a $4$-component spinor state \cite{beenakker_annihilation_2014}:

\begin{equation}
c(\xi)= 
\left( 
\begin{matrix}
c_{e, \uparrow}(\xi) \\
c_{e,\downarrow}(\xi)\\
c_{h,\uparrow}(\xi)\\
c_{h, \downarrow}(\xi)\\
\end{matrix}
\right),
\label{spinor}
\end{equation}
where $e$ ($h$) indicates the electron (hole) state and $\uparrow$ ($\downarrow$) the up (down) spin direction and $\xi$ is the energy of the incoming excitation. The particle-hole symmetry in Eq. (\ref{particle_hole}) leads to the constraint
$
 c(\xi)= \tau_{x}\otimes I \left[ c^{\dagger}(-\xi)\right]^{T}
$  \cite{alice_new_2012}.

The parameters which enter the transfer matrix are: $\delta=W/v_{F}$ the time required for the excitation to cross the SC region; 
$\alpha\approx W/l_{s}$ ($l_{s}=\hbar v/\Delta_{s}$ the proximity-induced coherence length) and $\beta \approx W/l_{m}$ with $l_{m}=(\hbar /e B)^{\frac{1}{2}}$ the magnetic length of the Hall system ($B$ the magnetic field); $\phi$ the phase of the SC; $\gamma$ and $\gamma'$ the relative phase shifts of electrons and holes in presence of the magnetic field \cite{beenakker_annihilation_2014,ostaay_spin_2011,hoppe_andreev_2000}.

After some algebra, the explicit form of this transfer matrix which takes into account Andreev reflection reads \cite{beenakker_annihilation_2014}:
\begin{equation}
\mathcal{M}(\xi)= e^{i \xi \delta} e^{i \Gamma \tau_{z}} \mathcal{U}(\tilde{\theta}, \phi) e^{i \Gamma' \tau_{z}}
\label{transfer_exp}
\end{equation}
where we introduced the phase shifts 
$\Gamma= \gamma+ \Omega$, $\Gamma'= \gamma'+ \Omega$, with $\Omega= \arctan\left(\beta \tan{\sqrt{\alpha^{2}+\beta^{2}}}/\sqrt{\alpha^{2}+\beta^{2}} \right)$ and the matrix
\begin{equation}
\mathcal{U}(\tilde{\theta}, \phi) =\exp{\left[i \tilde{\theta} \sigma_{y} \otimes (\tau_{x} \cos{\phi}+ \tau_{y} \sin{\phi}) \right]},
\end{equation}
with $\tilde{\theta}$ an angle such that
\begin{equation}
\sin{\tilde{\theta}}= \alpha \sin(\sqrt{\alpha^{2}+\beta^{2}})/\sqrt{\alpha^{2}+\beta^{2}}.
\end{equation}

By comparing the expression for the upper critical magnetic field in a Type II SC in terms of the coherence length $B_{c}= \Phi_{0}/(2 \pi l^{2}_{s})$ and the definition of the magnetic length $l_{m}$, the condition $l_{s}\ll l_{m}$ (and consequently $\alpha \gg \beta$) must be enforced to preserve the superconductivity. 

We consider now a SES injecting a spin up electron with the same WP as in Eq. (\ref{emitted_wp}) into the SC region. For the sake of simplicity we will investigate the zero temperature case. The outgoing state from the SC is a Bogoliubov quasiparticle (a coherent electron/hole superposition  with opposite spin):
\begin{eqnarray}
|\mathcal{B}\rangle&=&\mathcal{W}_{e} |e, \uparrow \rangle+ \mathcal{W}_{h} |h, \downarrow \rangle\nonumber\\
&=&\cos{\tilde{\theta}} |e, \uparrow \rangle+ \sin{\tilde{\theta}}e^{-i\Phi} |h, \downarrow \rangle
\label{Bogoliubov}
\end{eqnarray}
with $\Phi=2 \Gamma- \phi$ and $|e, \uparrow \rangle$, $|h, \downarrow \rangle$ a notation for electron/hole outgoing states from the SC region.

The averaged total current and particle density outgoing from the SC region are defined as:
\begin{eqnarray}
\langle \varphi |I |\varphi \rangle &=&-e \langle \varphi |:\tilde{\Psi}^{\dagger} \tau_{z} \tilde{\Psi}: |\varphi 
\label{current_def}\rangle\\
\langle \varphi |\rho  |\varphi \rangle &=& \langle \varphi |:\tilde{\Psi}^{\dagger} \tilde{\Psi}: |\varphi \rangle
\label{density}
\end{eqnarray}
where $-e<0$ and we have omitted the time dependence for notational convenience. Note that, in the above expressions, the definition 
$\tilde{\Psi}(t)=(4\pi)^{1/2} \int^{+\infty}_{-\infty} d \xi e^{-i \xi t} \mathcal{M}(\xi) c(\xi)$ 
for the outgoing spinor is required to avoid double counting \cite{beenakker_annihilation_2014}.

Applying Wick's theorem and considering well localized WPs in the positive energy domain, the current reduces to: 
\begin{eqnarray}
\langle \varphi |I(t) |\varphi \rangle
=-e\cos(2\tilde{\theta}) \varphi(t- \delta) \varphi^{*}(t- \delta).
\label{current}
\end{eqnarray} 
The outgoing electronic current of Eq. (\ref{current}) differs from the incoming one \cite{grenier_electron_2011,ferraro_wigner_2013} $\langle \varphi |I_{in}(t) |\varphi \rangle \equiv -e \varphi(t) \varphi^{*}(t)$, by a time delay $\delta$ and by a factor $\cos(2\tilde{\theta})$, which takes into account the conversion of electrons into holes via Andreev reflections. This is simply the difference between the probability $|\mathcal{W}_{e}|^{2}= \cos^{2}{\tilde{\theta}}$ for the incoming electron to emerge as an electron and $|\mathcal{W}_{h}|^{2} =\sin^{2}{\tilde{\theta}}$ to be converted into a hole. For $\tilde{\theta}=0$ ($|\mathcal{W}_{e}|^{2}=1$ and $|\mathcal{W}_{h}|^{2}=0$) the SC contact only induces a delay, while for $\tilde{\theta}=\pi/2$ ($|\mathcal{W}_{e}|^{2}=0$ and $|\mathcal{W}_{h}|^{2}=1$) the incoming electron is completely converted into a hole and a Cooper pair enters into the SC. More importantly, for $\tilde{\theta}=\pi/4$ ($|\mathcal{W}_{e}|^{2}=|\mathcal{W}_{h}|^{2}=1/2$) the electron and hole contributions compensate and no averaged current flows out. Nevertheless, this zero averaged current still bears fluctuations. The charge outgoing from the SC contact,  
\begin{equation}
\mathcal{Q}= \int^{+\infty}_{-\infty} dt \langle \varphi |I(t) |\varphi \rangle= -e \cos(2\tilde{\theta})
\label{charge}
\end{equation}
is not conserved as a consequence of the creation/destruction of Cooper pairs in the SC. Conversely, due to the unitarity of the scattering matrix, the outgoing excitation density is given by: 
\begin{eqnarray}
\langle \varphi |\rho(t) |\varphi \rangle= \varphi (t- \delta) \varphi^{*}(t- \delta)= \langle \varphi |\rho_{in}(t-\delta) |\varphi \rangle,\nonumber\\
\end{eqnarray}
which implies a mere time delay $\delta$ with respect to the incoming one. The prefactor is given by $|\mathcal{W}_{e}|^{2}+|\mathcal{W}_{h}|^{2}=1$, which illustrates the conservation of the number of injected excitations: 
\begin{equation}
\mathcal{N}= \int^{+\infty}_{-\infty} dt \langle \varphi |\rho_{in}(t-\delta) |\varphi \rangle=1~.
\end{equation}
Both the non-conservation of the charge and the conservation of the excitation number are encoded in the Bogoliubov-de Gennes Hamiltonian.

We also consider the current noise at the output of the SC contact: 
\begin{eqnarray}
\mathcal{S}_{source}&=&\int^{+\infty}_{-\infty} dt dt' \langle \varphi| I(t) I(t')  |\varphi\rangle_{c} \nonumber\\
&=& e^{2}\sin^{2}(2\tilde{\theta}).
\label{noise_out}
\end{eqnarray}

The above quantity is proportional to $P_{e} \cdot P_{h}$: it vanishes in the absence of a SC contact ($\tilde{\theta}=0$), as expected 
\cite{mahe_current_2010,bocquillon_coherence_2013}, and when the incoming electron is completely converted into a hole ($\tilde{\theta}=\pi/2$). Its maximum is reached for $\tilde{\theta}=\pi/4$, when the outgoing averaged current is zero. 

\subsection{Cross-correlated noise in a QPC geometry.}\label{Noise}
\begin{figure}[h]
\centering
\includegraphics[scale=0.35]{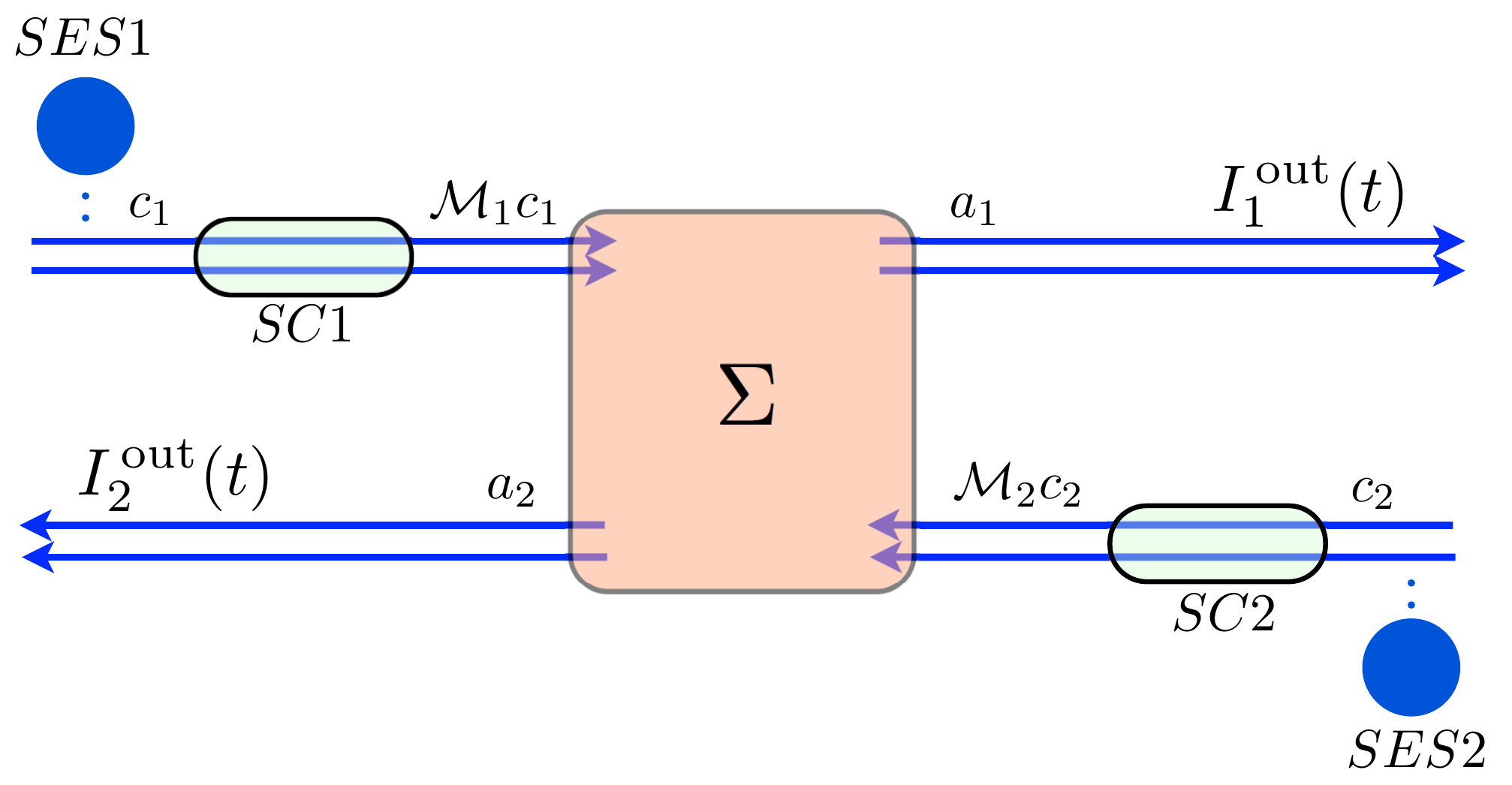}
\caption{(Color online) QPC geometry for Bogoliubov quasiparticles. The individual electron sources $SES1$ and $SES2$ inject electrons (described by the operators $c_{j}$ ($j=1,2$)), into the two SC contacts $SC1$ and $SC2$. The outgoing excitation ($\mathcal{M}_{j}c_{j}$) reach the QPC and are partitioned according to the scattering matrix $\Sigma$. We consider the cross-correlated noise between the outgoing currents $I^{\,\mathrm{out}}_{j}(t)$ (written in terms of the operators $a_{j}$). Picture taken from \cite{ferraro_nonlocal_2015}.}
\label{fig9}
\end{figure}

We now investigate the outgoing cross-correlated noise in a QPC geometry where one or two SES ($SES1$ and $SES2$) inject electronic WPs with spin up in the vicinity of one or two SC regions (see Fig. \ref{fig9}). 

The annihilation spinors outgoing from the QPC are related to the ones emitted by the two SES through the scattering matrix, which is parametrized by transmission (reflection) coefficients $\mathcal{T}$ ($\mathcal{R}$).
The zero frequency cross-correlated noise outgoing from the QPC is again given by Eq. (\ref{eq:noisestart}). 

\subsubsection{Hanbury-Brown-Twiss contribution.}\label{Section_HBT}
When only one of the two SES (labeled $j$) is ``on'' we obtain the HBT contribution to the noise. At zero temperature, the injected excitations crossing the SC contact are converted into Bogoliubov quasiparticles which reach the QPC and get partitioned \cite{bocquillon_electron_2012a}. This contribution to the noise is:
\begin{equation} 
\mathcal{S}_{HBT}^{(j)}
=  -e^{2} {\mathcal R}\mathcal{T} \cos^{2}(2\tilde{\theta}_{j}).
\label{HBT}
\end{equation}

This represents the shot noise associated with a WP carrying charge $\mathcal{Q}$ (see Eq. (\ref{charge})) and is therefore proportional to $\mathcal{Q}^{2}$. In the absence of SC ($\tilde{\theta}=0$), $\mathcal{S}_{HBT}=-e^{2} {\mathcal R \mathcal{T}}$ as expected (see Eq. (\ref{eq:SHBT}) in the zero temperature limit). For $\tilde{\theta}_{j}=\pi/4$, the state which reaches the QPC (a balanced superposition of electrons and holes) generates no noise at all. This is because this individual zero charged quasiparticle excitation presents a non trivial internal structure which cannot be simply described in terms of an incoherent mixture of electrons and holes \cite{moskalets_dissipation_2002,rychkov_photon_2005}. For $\tilde{\theta}_{j}=\pi/2$, the electron is completely converted into a hole, and the noise is the same as for $\tilde{\theta}_{j}=0$.   

\subsubsection{Hong-Ou-Mandel contribution.}\label{Section_HOM}
If both the SES are ``on'' we obtain the HOM noise signal \cite{jonckheere_electron_2012,bocquillon_coherence_2013}
\begin{equation}
\mathcal{S}_{HOM}= \Delta \mathcal{S}_{HOM}+ \mathcal{S}_{HBT}^{(1)}+\mathcal{S}_{HBT}^{(2)}
\end{equation}
with
\begin{eqnarray}
\Delta \mathcal{S}_{HOM}/\mathcal{S}_{0}&=& A(\delta_{1}-\delta_{2}-\eta)\left[ 1+ \cos(2\tilde{\theta}_{1})\cos(2\tilde{\theta}_{2})\right.\nonumber\\
&&~\left. -\cos(\Phi_{12}) \sin(2\tilde{\theta}_{1})\sin(2 \tilde{\theta}_{2})\right],
\label{HOM_general}
\end{eqnarray}
$\mathcal{S}_{0}= e^{2} {\mathcal R}\mathcal{T}$, $\eta$ the time delay in the emission between the two SES, $\Phi_{jk}= 2 \Gamma_{j}- 2\Gamma_{k}-\phi_{j}+\phi_{k}$, and the squared overlap 
$A(\tau)=\left|\int^{+\infty}_{-\infty} dt  \varphi^{*}(t- \tau) \varphi(t)\right|^2$ between identical WPs with a delay $\tau$.
This constitutes a general, central analytical result, as it addresses the HOM collision of two unsynchronized Bogoliubov quasiparticles.

When the two SC regions only differ in their order parameter phase and the two SES are properly synchronized ($\eta=0$) one obtains the simplified expression: 
\begin{equation}
\mathcal{S}_{HOM}^{2SC}= e^{2}{\mathcal R}\mathcal{T}\sin^{2}(2\tilde{\theta}) \left[1-\cos(\phi_{1}-\phi_{2}) \right]
\label{HOM_2SC}
\end{equation}
which clearly shows a non-local dependence on the difference of the SC order parameter phases as already pointed out in Ref. \cite{beenakker_annihilation_2014} in the DC regime. The device shows no dependence on the SC phase at the level of the averaged current (first order coherence), but presents an oscillatory modulation in the noise (second order coherence). This is a clear demonstration of the fact that noise measurements in this kind of devices allow to access purely two-quasiparticle effect in analogy to what was discussed in the framework of IQH effect for the interferometers of Refs.\cite{samuelsson_two_2004,neder_interference_2007} or the revisitation of the Franson interferometer \cite{franson_bell_1989} proposed in Refs. \cite{splettstoesser_two_2009,thibierge_two_2016}.

If in Eq. (\ref{HOM_2SC}) $\phi_1-\phi_2\neq 0$ (mod. $2\pi$), the noise vanishes only when two electrons or two holes reach the QPC at the same time ($\tilde{\theta}=0$ or $\tilde{\theta}=\pi/2$) as a consequence of the Pauli principle \cite{jonckheere_electron_2012}. Remarkably, the noise reaches its maximum for $\tilde{\theta}= \pi/4$. To explain this, look at the structure of the $\Delta \mathcal{S}_{HOM}$ term in Eq. (\ref{HOM_general}). By considering two Bogoliubov excitations in the form of Eq. (\ref{Bogoliubov}) simultaneously reaching the QPC, this contribution to the noise is proportional to: 
\begin{eqnarray}
&&|\mathcal{W}^{1}_{e} {\mathcal{W}^{2}_{e}}^{*}-\mathcal{W}^{1}_{h} {\mathcal{W}^{2}_{h}}^{*}|^{2}=\nonumber\\
&& | \cos{\tilde{\theta}_{1}} \cos{\tilde{\theta}_{2}}-\sin{\tilde{\theta}_{1}} \sin{\tilde{\theta}_{2}} e^{-i( \Phi_{1}-\Phi_{2})}|^{2}.
\end{eqnarray} 
It thus corresponds to the difference between the product of electron and hole probability amplitudes. In particular for $\tilde{\theta}_{1}=\tilde{\theta}_{2}=\pi/4$ the Bogoliubov quasiparticles carry zero charge and zero shot noise, but are by far not trivial excitations with a complex structure given by the coherent superposition of electrons and holes which can be detected only at the level of the two quasiparticle interferometry. The above argument is also useful in order to understand why the HOM contribution to the noise in Eq. (\ref{HOM_2SC}) is zero for $\phi_1-\phi_2= 0$ (mod. $2\pi$). Under this condition indeed, the $\Delta \mathcal{S}_{HOM}$ term exactly compensates the two (equal) $\mathcal{S}_{HBT}$ contributions in analogy to what occurs for the IQH case in absence of interaction.

\begin{figure}[h]
\centering
\includegraphics[scale=0.4]{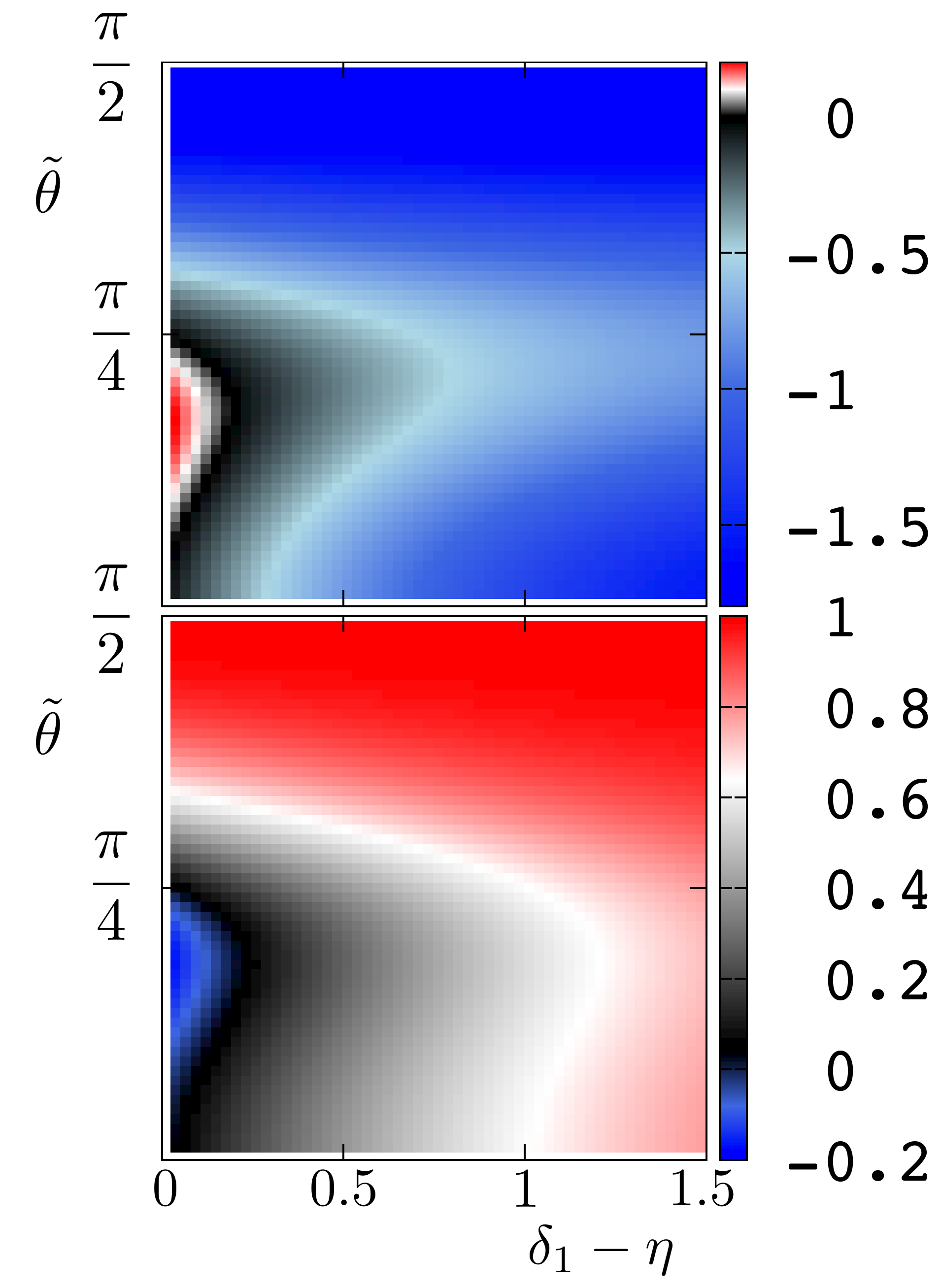}
\caption{(Color online). Upper panel. Density plot of $\mathcal{S}_{HOM}^{1SC}$ in units of $\mathcal{S}_{0}=e^{2}\mathcal{R}\mathcal{T}$. ``Blue'' identifies negative regions where the hole contribution dominates over the electron one, while the little red area represents the positive noise. Lower panel. Density plot of $R^{1SC}$ as a function of $\delta_{1}-\eta$ and $\tilde{\theta}$. The blue area corresponds to negative values of the ratio which cannot be reached in conventional electronic quantum optics experiments. Picture taken from \cite{ferraro_nonlocal_2015}.}
\label{fig10}
\end{figure}

The peculiar structure of the HOM noise contribution for two synchronized Bogoliubov excitations directly reflects into the divergences associated with the ratio:
\begin{eqnarray}
R^{2SC}&=& \frac{\mathcal{S}_{HOM}^{2SC}}{\mathcal{S}_{HBT}^{(1)}+\mathcal{S}_{HBT}^{(2)}}\nonumber\\
&=& -\frac{1}{2} \tan^{2}(2 \tilde{\theta})  \left[1-\cos(\phi_{1}-\phi_{2}) \right].
\end{eqnarray}
We can also achieve collisions between single electrons and Bogoliubov quasiparticles: it is now shown that the electron colliding with the Bogoliubov quasiparticle allows to probe the content of the latter. Starting from Eq. (\ref{HOM_general}) the corresponding noise becomes:
\begin{eqnarray}
\mathcal{S}_{HOM}^{1SC}&=& e^{2}{\mathcal R}\mathcal{T} 
\left\{\left[ 1+\cos(2\tilde{\theta})\right]A(\delta_{1}-\eta)\right.\nonumber \\
&~&- \left.\cos^{2}(2\tilde{\theta})-1\right\}
\end{eqnarray}
with a maximum WP overlap $A=1$, the reference electron interferes with: a) another electron ($\tilde{\theta}=0$) leading to a zero noise (Pauli principle); b) with a hole ($\tilde{\theta}=\pi/2$) with a consequent minimum of the noise \cite{jonckheere_electron_2012}; c) a more general Bogoliubov quasiparticle. In the latter case the cross-correlated noise assumes positive values when the electron component of the Bogoliubov excitation dominates over the hole one ($|\mathcal{W}_{e}|^{2}>|\mathcal{W}_{e}|^{2}$ and consequently for $0<\tilde{\theta}<\pi/4$). By decreasing the WP overlap, the $\Delta \mathcal{S}_{HOM}$ contribution to the noise is suppressed. However, even away from perfect synchronization, it is possible to observe positive and negative regions from which we can extract the dominant contribution to the Bogoliubov quasiparticle. The situation in the case of a WP exponential in time (see Eq. (\ref{emitted_wp})), where $A(\tau)= e^{-\Gamma |\tau|}$, is illustrated by the density plot in the upper panel of Fig. \ref{fig10}.   

This represents an extremely useful tool to extract information about the structure of the Bogoliubov excitations through interferometric experiments with a known source (electronic WP). These observations indicate that the considered setup offers richer possibilities to implement a tomographic protocol by means of HOM interferometry with respect to what was proposed in the electronic case \cite{ferraro_real_2014}.
Another relevant quantity to explore is given by the ratio: 
\begin{eqnarray}
R^{1SC}&=&  \mathcal{S}^{1SC}_{HOM}/(\mathcal{S}^{(1)}_{HBT}+\mathcal{S}^{(2)}_{HBT})\nonumber\\
 &=&1- \frac{1+ \cos(2 \tilde{\theta})}{1+ \cos^{2}(2 \tilde{\theta})}A(\delta_{1}-\eta) .
\end{eqnarray}
This also contains negative regions (blue areas in the lower panel of Fig. \ref{fig10}) which are forbidden in conventional electron quantum optics situations ($\tilde{\theta}=0$) \cite{jonckheere_electron_2012} due to the constraints imposed by the charge conservation.

\section{Conclusions.}\label{Conclusions}

Scattering theory thus allows to study a number of EQO interferometric situations. In IQH regime, theory has already been confronted with experiment \cite{bocquillon_coherence_2013}. The prediction of the HOM dip is solid, but its visibility does not correspond to what was expected \cite{jonckheere_electron_2012}. Experiments cannot easily be achieved at filling factor one  and when several channels are present, interactions between them lead to charge fractionalization thus decoherence effects accompanied by a reduced visibility \cite{wahl_interactions_2014}. Two new directions for EQO have been proposed: the study of HOM collisions in the QSH effect in 2D topological insulators \cite{ferraro_electronic_2014} and the possibility of observing non-local phenomena when colliding two Bogoliubov quasiparticles at the location of a QPC \cite{ferraro_nonlocal_2015}. 

In these two setups, finite temperature effects were discarded. This should not change dramatically our predictions as long as the electrons WPs have a well defined energy above the Fermi sea and, for the former case, well below the gap. However they are likely to be affected when considering electron-hole interferometry \cite{jonckheere_electron_2012} where the overlap between electron and hole WPs within an energy window set by the temperature is crucial. 

Further work should however address the issue of interchannel interactions as in Ref. \cite{wahl_interactions_2014}. For the QSH setup, one suspects that Coulomb interactions between the counter-propagating edges could lead to a further decreasing in the visibilities of the \emph{Pauli} and the $\mathcal{Z}_{2}$ dips  in the two electron collisions, and prevent a maximal visibility for the three electron dip. For Ref. \cite{ferraro_nonlocal_2015}, the approach of Ref. \cite{wahl_interactions_2014} could be directly transposed in order to predict the modification of the Bogoliubov quasiparticle impinging on the QPC, and the modification of the HOM signal. 

Finally, EQO could also be studied in intrinsically interacting systems, which represent a strongly correlated state of matter, such as the fractional quantum Hall effect, in order to study HBT and  HOM interferometry of electrons/Laughlin quasiparticles. In this direction, the characterization of a single quasiparticle source was recently proposed \cite{ferraro_single_2015}, and the HBT setup was studied in Ref. \cite{rech_minimal_2016} for Lorentzian WPs.    

\begin{acknowledgement}
The support of Grant No. ANR-2010-BLANC-0412 (``1 shot'') and of ANR-2014-BLANC ``one shot reloaded'' is acknowledged. This work was carried out
in the framework of Labex ARCHIMEDE Grant No. ANR-11-LABX-0033 and of A*MIDEX project Grant No. ANR-
11-IDEX-0001-02, funded by the ``investissements d'avenir'' French Government program managed by the French National
Research Agency (ANR).
\end{acknowledgement}

%
%
\bibliography{biball_2016b}{}

\end{document}